\documentclass[a4paper,11pt]{article}
\pdfoutput=1 

\usepackage{jcappub} 

\usepackage[T1]{fontenc} 
\usepackage{verbatim}
\usepackage[utf8]{inputenc}
\usepackage[american]{babel}
\usepackage{graphicx}
\usepackage{epsfig}
\usepackage{adjustbox}
\usepackage{booktabs}
\usepackage{multirow}
\usepackage{dcolumn}
\usepackage{amsmath}
\usepackage{amsfonts}
\usepackage{mathtools}
\usepackage{amssymb}
\usepackage{empheq}
\usepackage{subcaption} 
\usepackage{caption}
\usepackage{epstopdf}
\usepackage{bm}
\usepackage{bbm}
\usepackage{dsfont}
\usepackage{mathrsfs}
\usepackage{version}
\usepackage{enumerate}
\usepackage{braket}
\usepackage{enumitem}
\usepackage{transparent}
\usepackage{pifont}
\usepackage{colortbl}
\usepackage{rotating}
\usepackage{adjustbox}
\usepackage{cancel}
\usepackage{tabularx}
\usepackage{soul}
\usepackage{pdfpages}
\usepackage{microtype}
\usepackage{ulem}
\usepackage[table]{xcolor}
\usepackage{siunitx}
\usepackage{float} 
\usepackage[nocompress]{cite} 
\usepackage{marginnote}
\usepackage{relsize}
\usepackage{feynmp-auto}

\usepackage{hyperref}
\hypersetup{colorlinks=true,
linkcolor=navyblue,citecolor=coralred,urlcolor=green(munsell),pdfencoding=auto}

\definecolor{navyblue}{rgb}{0.0, 0.0, 0.5}
\definecolor{bleudefrance}{rgb}{0.19, 0.55, 0.91}
\definecolor{coralred}{rgb}{1.0, 0.25, 0.25}
\definecolor{royalblue}{rgb}{0.25, 0.41, 0.88}
\definecolor{cadmiumgreen}{rgb}{0.0, 0.42, 0.24}
\definecolor{green(munsell)}{rgb}{0.0, 0.66, 0.47}
\definecolor{blue-violet}{rgb}{0.54, 0.17, 0.89}
\definecolor{darkviolet}{rgb}{0.58, 0.0, 0.83}
\definecolor{orange(colorwheel)}{rgb}{1.0, 0.5, 0.0}
\definecolor{internationalorange}{rgb}{1.0, 0.31, 0.0}

\definecolor{magenta(process)}{rgb}{1.0, 0.0, 0.56}

\definecolor{darkspringgreen}{rgb}{0.09, 0.45, 0.27}

\definecolor{royalblue(web)}{rgb}{0.25, 0.41, 0.88}

\definecolor{cadmiumorange}{rgb}{0.93, 0.53, 0.18}

\definecolor{heliotrope}{rgb}{0.87, 0.45, 1.0}

\makeatletter
\renewcommand*{\@textcolor}[3]{%
\protect\leavevmode
\begingroup
\color#1{#2}#3%
\endgroup
}
\makeatother

\newcommand{\myfloatalign}{\centering}

\newcommand{\hp}{\hphantom}
\newcommand\sm[1]{$\smash{#1}$}

\makeatletter
\newlength{\apb@width}
\newcommand{\autoparbox}[2][c]{\settowidth{\apb@width}{#2}\parbox[#1]{\apb@width}{#2}}

\makeatother

\DeclareCaptionLabelSeparator{colquad}{.\,\,\,}
\renewcommand\[{\left[}

\makeatletter
\let\save@mathaccent\mathaccent
\newcommand*\if@single[3]{%
\setbox0\hbox{${\mathaccent"0362{#1}}^H$}%
\setbox2\hbox{${\mathaccent"0362{\kern0pt#1}}^H$}%
\ifdim\ht0=\ht2 #3\else #2\fi
}
\newcommand*\rel@kern[1]{\kern#1\dimexpr\macc@kerna}
\newcommand*\widebar[1]{\@ifnextchar^{{\wide@bar{#1}{0}}}{\wide@bar{#1}{1}}}
\newcommand*\wide@bar[2]{\if@single{#1}{\wide@bar@{#1}{#2}{1}}{\wide@bar@{#1}{#2}{2}}}
\newcommand*\wide@bar@[3]{%
\begingroup
\def\mathaccent##1##2{%
\let\mathaccent\save@mathaccent
\if#32 \let\macc@nucleus\first@char \fi
\setbox\z@\hbox{$\macc@style{\macc@nucleus}_{}$}%
\setbox\tw@\hbox{$\macc@style{\macc@nucleus}{}_{}$}%
\dimen@\wd\tw@
\advance\dimen@-\wd\z@
\divide\dimen@ 3
\@tempdima\wd\tw@
\advance\@tempdima-\scriptspace
\divide\@tempdima 10
\advance\dimen@-\@tempdima
\ifdim\dimen@>\z@ \dimen@0pt\fi
\rel@kern{0.6}\kern-\dimen@
\if#31
\overline{\rel@kern{-0.6}\kern\dimen@\macc@nucleus\rel@kern{0.4}\kern\dimen@}%
\advance\dimen@0.4\dimexpr\macc@kerna
\let\final@kern#2%
\ifdim\dimen@<\z@ \let\final@kern1\fi
\if\final@kern1 \kern-\dimen@\fi
\else
\overline{\rel@kern{-0.6}\kern\dimen@#1}%
\fi
}%
\macc@depth\@ne
\let\math@bgroup\@empty \let\math@egroup\macc@set@skewchar
\mathsurround\z@ \frozen@everymath{\mathgroup\macc@group\relax}%
\macc@set@skewchar\relax
\let\mathaccentV\macc@nested@a
\if#31
\macc@nested@a\relax111{#1}%
\else
\def\gobble@till@marker##1\endmarker{}%
\futurelet\first@char\gobble@till@marker#1\endmarker
\ifcat\noexpand\first@char A\else
\def\first@char{}%
\fi
\macc@nested@a\relax111{\first@char}%
\fi
\endgroup
}
\makeatother

\newcommand\ee{\end{equation}}
\newcommand\be{\begin{equation}}
\newcommand\eea{\end{eqnarray}}
\newcommand\bea{\begin{eqnarray}}
\newcommand{\bsp}{\begin{split}}
\newcommand{\esp}{\end{split}}
\newcommand{\bit}{\begin{itemize}[leftmargin=*]}
\newcommand{\eit}{\end{itemize}}
\newcommand{\ben}{\begin{enumerate}[leftmargin=*]}
\newcommand{\een}{\end{enumerate}}

\renewcommand{\emph}{\textit}
\newcommand{\cmark}{\checkmark}
\newcommand{\xmark}{\ding{55}}

\newcommand\eq[1]{Eq.~\eqref{eq:#1}}
\newcommand\eqsI[1]{Eqs.~\eqref{eq:#1}}
\newcommand{\eqsII}[2]{Eqs.~\eqref{eq:#1}, \eqref{eq:#2}}

\newcommand{\iu}{\mathrm{i}}
\newcommand{\eu}{\mathrm{e}}
\newcommand{\dif}{\mathrm{d}}

\renewcommand{\vec}{\bm}

\def\tr{{{^{(3)}}\!R}}

\def\mpl{M_{\rm P}}

\def\<{\left\langle}
\def\>{\right\rangle}

\def\comment#1{}

\title{Zoology of Graviton non-Gaussianities}

\author[a]{Giovanni Cabass}

\affiliation[a]{School of Natural Sciences, Institute for Advanced Study, 1 Einstein Dr, Princeton, NJ 08540}

\emailAdd{gcabass@ias.edu}

\abstract{\noindent We characterize graviton non-Gaussianities in models of inflation where de Sitter 
boosts are spontaneously broken. We discuss which of the symmetry breaking patterns studied in Nicolis et al., 2015 
\cite{Nicolis:2015sra} can sustain a period of quasi de Sitter expansion, and show that the 
symmetry breaking pattern of Solid Inflation allows the most freedom for graviton interactions. 
We comment on the phenomenological consequences of some of these interactions. As a byproduct of 
this analysis we construct the ``EFT of Solid Inflation'', which has all the useful features 
of the EFT of Inflation but for the case of broken spatial diffeomorphisms.}

\begin{document}
\maketitle
\flushbottom



\section{Introduction} 
\label{sec:introduction}

\noindent Upcoming CMB experiments will target the tensor-to-scalar ratio \sm{r} to a sensitivity 
of \sm{\sigma_r\sim 10^{-3}} \cite{Matsumura:2013aja,Abazajian:2016yjj,Finelli:2016cyd,Delabrouille:2017rct,
Ade:2018sbj,Hanany:2019lle,Shandera:2019ufi,Abazajian:2019eic,Hazumi:2021yqq}. 
If primordial \sm{B} modes are detected, the way towards constraints on tensor non-Gaussianities, 
and then on the interactions of the graviton during inflation, will open. 

Even in the less exciting case that these experiments will only improve the current upper bounds, 
it is still very interesting to look at what are the possible interactions in quasi de Sitter backgrounds. This is because of recent developments in the ``Cosmological Bootstrap'' 
program \cite{Maldacena:2011nz,Mata:2012bx,Ghosh:2014kba,Pajer:2016ieg,
Arkani-Hamed:2018kmz,Green:2020ebl,Baumann:2020dch,Pajer:2020wnj,Goodhew:2020hob,Cespedes:2020xqq,Pajer:2020wxk}. Indeed, Ref.~\cite{Pajer:2020wxk} has 
discussed how to extend the bootstrap rules to the scenario where de Sitter boosts are broken, more 
precisely to the phenomenologically-interesting case of single-clock inflation. 
Working at the level of the Lagrangian can help in developing the bootstrap rules, 
and also identify regions of parameter space where non-Gaussianities can be enhanced, 
as is the case for the scalar non-Gaussianity in \sm{P(X,\phi)} theories. 

The case of single-clock inflation is particularly simple to study since only one Goldstone mode is present. 
Once we move away from this symmetry breaking pattern the number of Goldstone modes increases, and a study 
of all the possible non-Gaussianities (scalar, vector and tensor) becomes complicated. Since restricting to graviton interactions 
still allows to highlight the new features of the other symmetry breaking patterns, we have decided to study these first. 

Graviton non-Gaussianities in single-clock inflation have been investigated in detail in \cite{Maldacena:2002vr,Maldacena:2011nz}. 
What happens when couplings with the foliation are included has been discussed in \cite{Creminelli:2014wna,Bordin:2017hal,Bordin:2020eui,Bartolo:2020gsh}. 
How do we extend the analysis to different symmetry breaking patterns? 
Ref.~\cite{Nicolis:2015sra} provides a guide for how to achieve this. 
On scales shorter than the horizon the de Sitter isometries reduce to the Poincar{\' e} group 
(special conformal transformations and ``de Sitter dilations'' -- the combination of time translations with spatial dilations -- reduce to 
Lorentz boosts and time translations, respectively). Then, we can gain intuition from the flat-space analysis 
of \cite{Nicolis:2015sra}, in which the various ways to break spacetime symmetries while preserving spatial rotations and translations (both space and time) have been classified. 

The structure of this paper is as follows. In Section~\ref{sec:zoology_review} 
we review the results of \cite{Nicolis:2015sra} and discuss the coupling with gravity. 
In Section~\ref{sec:inflation_and_actions} we discuss which of 
the different symmetry breaking patterns admit a quasi de Sitter 
solution once coupled to gravity, and consequently build the unitary-gauge 
action. The main section of the paper is, then, Section~\ref{sec:main_section}: there we study 
what are the graviton interactions in the different scenarios and show that the symmetry breaking pattern 
of solids, of which Solid Inflation \cite{Endlich:2012pz} is the lowest order in a derivative expansion, 
is the one that allows the most freedom. 
In this section we also discuss some phenomenological consequences of these interactions, and make a comparison 
to the symmetry breaking pattern of a superfluid. For simplicity, when 
discussing phenomenology we focus on quadratic and cubic graviton self-couplings up to cubic order in derivatives: 
for each of these we list all the corresponding operators in the unitary-gauge action. 
We conclude in Section~\ref{sec:conclusions}. 

Appendices~{\ref{app:appendix-new},} \ref{app:appendix-A}{,} \ref{app:appendix-B} {and \ref{app:appendix-C}} 
contain some details of the calculations carried out in Sections~\ref{sec:inflation_and_actions} 
and \ref{sec:main_section}. 
More precisely, Appendi{ces}~\ref{app:appendix-new} {and \ref{app:appendix-A}} show 
how to construct the equivalent of the Effective Field Theory (EFT) of Inflation \cite{Cheung:2007st} 
for the case of broken spatial diffeomorphisms, with special emphasis on the property of ``tadpole cancellation'' 
(i.e.~the fact that we want to write an EFT for fluctuations around a FLRW metric). 

\paragraph{Summary of main results} We summarize the main results of the paper in the compact table below. 
First, we list the different ways to break de Sitter boosts and what they correspond to once we make gravity dynamical. For each of them 
we show what are the available (\cmark) building blocks in terms of the transverse and traceless graviton field \sm{\gamma_{ij}}. 
These building blocks can be combined by contracting indices in an \sm{SO(3)}-invariant way. 
Given that for all the symmetry breaking patterns it is possible to take further time and spatial derivatives of the available building 
blocks, there is no loss of generality in stopping at the order in derivatives we show in the table.

\begin{table}[H]
\myfloatalign
\label{tab:summary}
\centering
\medskip
\begin{tabular}{lccccc}
\toprule
{medium} & broken gauge symmetry & $\gamma_{ij}$ & $a^{-1}\partial_k\gamma_{ij}$ & $a^{-2}\partial_k\partial_l\gamma_{ij}$ & $\dot{\gamma}_{ij}$ \\
\midrule
superfluid & time diff.s & \xmark & \xmark & \cmark & \cmark \\[2ex]
type-I framid & local Lorentz boosts & \xmark & \xmark & \cmark & \cmark \\[2ex]
type-II framid & local boosts and rotations & \xmark & \cmark & \cmark & \cmark \\[2ex]
solid & spatial diff.s & \cmark & \cmark & \cmark & \cmark \\
\bottomrule
\end{tabular}
\end{table}

\section{\texorpdfstring{``Zoology of condensed matter'' and coupling with gravity}{"Zoology of condensed matter" and coupling with gravity}} 
\label{sec:zoology_review}

\noindent Let us review the analysis of \cite{Nicolis:2015sra}. We are interested in classifying all the symmetry breaking patterns 
that can be associated with a static, homogeneous, and isotropic medium in a relativistic theory. 

This implies that, together with the Poincar{\' e} generators \sm{P_\mu, K_i, J_i} of spacetime transformations, 
there is a set of translation and rotation generators \sm{\bar{P}_\mu, \bar{J}_i} that govern the excitations inside the medium and 
that leave the ground state invariant. These satisfy the same algebra as the Poincar{\' e} generators, i.e.~ 
\begin{equation}
\label{eq:zoology_review-1}
\big[\bar{J}_i,\bar{J}_j\big] = \iu\epsilon_{ijk}\bar{J}_k\,\,, \quad \big[\bar{J}_i,\bar{P}_j\big] = \iu\epsilon_{ijk}\bar{P}_k\,\,, 
\end{equation}
but they need not be the same as \sm{P_\mu, J_i}. Indeed, the different systems are classified by whether or not \sm{\bar{P}_\mu, \bar{J}_i} 
contain additional symmetries, generated by \sm{Q,Q_i,\tilde{Q}_i}: 
\begin{equation}
\label{eq:zoology_review-2}
\bar{P}_0 = P_0 + Q\,\,, \quad \bar{P}_i = P_i + Q_i\,\,, \quad \bar{J}_i = J_i + \tilde{Q}_i\,\,. 
\end{equation}

\subsection{The eightfold (sixfold) way} 
\label{subsec:sixfold}

\noindent Ref.~\cite{Nicolis:2015sra} identifies eight types of media, depending on which of the generators \sm{Q,Q_i,\tilde{Q}_i} are non-vanishing. However, 
only six of them can be realized by having these generators to be internal symmetries: 
the remaining two require them to not commute with the Poincar{\' e} generators. 
These two media are dubbed ``galileids'', since they are based on the galileon symmetry. Since, as the authors themselves argue, 
it is difficult to extend this symmetry to the case where gravity is dynamical, we will not consider them in the following. 
\begin{description}[leftmargin=0pt,labelwidth=0pt]
\item[Type-I framids\sm{\,\rm :} \quad \sm{\bar{P}_0 = P_0\,\,, \quad \bar{P}_i = P_i\,\,, \quad \bar{J}_i = J_i}] 
\leavevmode \\
\noindent This is the simplest scenario, since it does not involve any internal symmetry. It can be realized by 
having a vector field that acquires a vacuum expectation value \sm{\braket{A^\mu(x)} = \delta^\mu_0}. 
Ref.~\cite{Nicolis:2015sra} shows that there are three Goldstone modes present: they can be thought as the rapidity of a boost acting on \sm{\delta^\mu_0}. 
\item[Type-I superfluids\sm{\,\rm :} \quad \sm{\bar{P}_0 = P_0 + Q\,\,, \quad \bar{P}_i = P_i\,\,, \quad \bar{J}_i = J_i}] 
\leavevmode \\
\noindent The minimal setup to realize the superfluid scenario is to have a ``phase'' field \sm{\psi(x)} with vacuum 
expectation value \sm{\braket{\psi(x)} = t}. The Goldstone mode is the fluctuation of this phase field, \sm{\psi(x)=t+\pi(x)}. 
\item[Type-II framids\sm{\,\rm :} \quad \sm{\bar{P}_0 = P_0\,\,, \quad \bar{P}_i = P_i\,\,, \quad \bar{J}_i = J_i + \tilde{Q}_i}] 
\leavevmode \\
\noindent As we will see in the rest of the paper, this is an extension of the type-1 framid. It can be realized by a triplet of vector fields 
that rotate under an internal \sm{SO(3)} with vacuum expectation value \sm{\braket{A^\mu_i(x)} = \delta^\mu_i}. There are 
six Goldstone modes, which can be identified with the rapidity and Euler angles of boosts and rotations acting on \sm{\delta^\mu_i}. 
\item[Type-II superfluids\sm{\,\rm :} \quad \sm{\bar{P}_0 = P_0 + Q\,\,, \quad \bar{P}_i = P_i\,\,, \quad \bar{J}_i = J_i + \tilde{Q}_i}] 
\leavevmode \\ 
\noindent This symmetry breaking pattern can be realized by a combination of a type-1 superfluid and a type-2 framid, with a total of seven Goldstone modes. 
\item[Solids and fluids\sm{\,\rm :} \quad \sm{\bar{P}_0 = P_0\,\,, \quad \bar{P}_i = P_i + Q_i\,\,, \quad \bar{J}_i = J_i}] 
\leavevmode \\
\noindent This is the case of isotropic solids (that is, solids with no preferred axes, or ``jellies''). The low-energy effective field theory 
can be characterized by a triplet of scalar fields \sm{\phi^i(x)} with vacuum expectation value 
\sm{\braket{\phi^i(x)} = x^i} (the Lagrangian and Eulerian coordinates of the volume elements of the solid, respectively). There 
are three Goldstone modes, \sm{\phi^i(x) = x^i+\pi^i(x)}. Fluids are obtained by imposing a symmetry under volume-preserving internal diffeomorphisms. 
\item[Supersolids\sm{\,\rm :} \quad \sm{\bar{P}_0 = P_0 + Q\,\,, \quad \bar{P}_i = P_i + Q_i\,\,, \quad \bar{J}_i = J_i + \tilde{Q}_i}] 
\leavevmode \\
\noindent Supersolids (with finite-temperature superfluids as a subset) are obtained by adding a type-I superfluid phase to the above case. 
\end{description} 
{It is important to emphasize that in this paper we consider the implementations that involve the minimal number 
of Goldstone modes. We discuss this in more detail in Section~\ref{subsec:a_word_on_Goldstone_modes} 
(one can also refer to \cite{Nicolis:2015sra} itself, more precisely their Section~2.1).}

\subsection{Coupling with gravity} 
\label{subsec:coupling_with_gravity}

\noindent Let us now discuss how to couple these systems with gravity. The precise derivation would involve 
the Callan-Coleman-Wess-Zumino (CCWZ) construction \cite{Coleman:1969sm,Callan:1969sn,Volkov:1973vd,Ivanov:1975zq}, also known as coset construction, 
similarly to what has been done in \cite{Delacretaz:2014oxa} for Einstein gravity. 
Here we take a much faster route: we just make an educated guess on the final result based on \cite{Delacretaz:2015edn,Bordin:2018pca}. 
Indeed, these papers employ the CCWZ construction to couple to gravity two of the patterns of the previous section. 
\begin{itemize}[leftmargin=*]
\item Ref.~\cite{Bordin:2018pca} showed that the coset construction for 
the type-I superfluid gives the EFT of Inflation \cite{Cheung:2007st}, in which time diffeomorphisms are broken. 
\item Ref.~\cite{Delacretaz:2015edn}, instead, showed that the type-I framid results in a breaking of 
local Lorentz boosts, intended as transformations in the tangent space: for example, 
given the vierbein \sm{e^\mu_A}, objects like \sm{\nabla_\mu e^\mu_0} can now appear in the action (\sm{\mu,\nu,\dots} and \sm{A,B,\dots} are diffeomorphism indices and local Lorentz ones, respectively). 
\end{itemize} 
How to couple the remaining symmetry breaking patterns to gravity is then clear. 
\begin{itemize}[leftmargin=*]
\item The case of a generic fluid, and more generally of a solid, amounts to the breaking of spatial diffeomorphisms, i.e.~to 
Solid Inflation \cite{Endlich:2012pz} at lowest order in a derivative expansion.\footnote{{We recall that the additional 
symmetries of a fluid, i.e.~the invariance under volume-preserving internal diffeomorphisms holds, prevent us from 
building a healthy inflationary model, as discussed at the end of Section~3 of \cite{Endlich:2012pz} (see also 
\cite{Endlich:2010hf,Gripaios:2014yha} for more discussions on the quantum mechanics of fluids). In this work, whose focus is on the interactions 
of the graviton, we will not be concerned with these issues and consider the quadratic action of the Goldstone modes to be 
that of a solid/``jelly''.}} 
\item Type-II framids also involve the breaking of local Lorentz transformations. While in the case of the type-I framid only \sm{e^\mu_0} is allowed, 
now the dreibein \sm{e^\mu_I} (\sm{I=1,2,3}) is what can appear in the action. 
\item In both these cases, the fact that we want spatial and internal rotations to be broken to the diagonal subgroup 
amounts to contracting the indices \sm{I,J,K,\dots} in an \sm{SO(3)}-invariant way. 
\item Finally, the finite-temperature superfluid results in the breaking of all diffeomorphisms, i.e.~in Supersolid Inflation at 
lowest order in derivatives \cite{Bartolo:2015qvr,Ricciardone:2016lym,Domenech:2017kno,Celoria:2020diz}, 
while type-II superfluids break time diffeomorphisms and local Lorentz transformations. 
\end{itemize}

\section{Inflationary solutions and unitary-gauge actions} 
\label{sec:inflation_and_actions}

\noindent We are now ready to construct the unitary-gauge action for the different symmetry breaking patterns. 
We focus on the cases of type-I superfluids, solids and framids only. As discussed at the end of the previous section, the 
remaining two patterns can be obtained straightforwardly from these four.

\subsection{Type-I superfluids} 
\label{subsec:EFTI}

\noindent The breaking of time diffeomorphisms results in the EFT of Inflation, for which we refer to the original paper \cite{Cheung:2007st}. 
The building blocks in this case are all diffeomorphism-invariant quantities constructed from the metric, together with all objects that can be 
constructed from \sm{\partial_\mu t}, or equivalently the normal \sm{n_\mu = -\partial_\mu t/\sqrt{-g^{00}}} to the hypersurfaces of constant clock. 

What is important to emphasize in this case is the following. The high degree of symmetry of the 
FLRW background ensures that is possible to obtain ``tadpole cancellation'' 
at all orders in derivatives. Let us consider the ``slow-roll action'', i.e.~\cite{Cheung:2007st} 
\begin{equation}
\label{eq:EFTI-general_action}
S = \int\dif^4x\,\sqrt{-g}\,\bigg\{\frac{\mpl^2}{2}R - c(t)g^{00} - \Lambda(t)\bigg\}\,\,. 
\end{equation} 
At any order in derivatives one can show that additional operators can be split in a perturbation and a background part, 
without introducing new degrees of freedom, in 
such a way that the background part is always reabsorbed by the free functions \sm{c} and \sm{\Lambda} (possibly after integration by parts). The chief example is that of the 
extrinsic curvature of constant-\sm{t} hypersurfaces, \sm{K_{\mu\nu} = \delta\!K_{\mu\nu} + H h_{\mu\nu}} (where \sm{h_{\mu\nu}=g_{\mu\nu}+n_\mu n_\nu}).

\subsection{Solids and fluids} 
\label{subsec:solids}

\noindent When breaking spatial diffeomorphisms, the building blocks are again all diffeomorphism-invariant 
quantities constructed from the metric, together with all objects constructed from \sm{\partial_\mu x^i} 
(see also \cite{Lin:2015cqa} for a discussion focused on the scalar and vector modes). 

At leading order in derivatives, the action is 
\begin{equation}
\label{eq:solids-general_action}
S = \int\dif^4x\,\sqrt{-g}\,\bigg\{\frac{\mpl^2}{2}R + F(X,Y,Z)\bigg\}\,\,, 
\end{equation} 
where 
\begin{equation}
\label{eq:solids-X_Y_Z}
X = g^{ii}\,\,, \quad Y = \frac{g^{ij}g^{ji}}{X^2}\,\,, \quad Z = \frac{g^{ij}g^{jk}g^{ki}}{X^3}\,\,. 
\end{equation} 
Choosing \sm{F} and \sm{\partial F/\partial X} appropriately, one can solve the equations of motion on the background 
for any value of the energy density and pressure. More precisely, quasi de Sitter expansion \sm{\varepsilon = {-\dot{H}/H^2}} 
can be achieved if 
\begin{equation}
\label{eq:solids-SR_condition}
\varepsilon = \frac{\partial\ln F}{\partial\ln X} \ll 1\,\,, 
\end{equation} 
and the role of the clock determining when inflation ends is played by \sm{X = 3/a^2(t)}. 

Compared to the EFT of Inflation, there are however some subtleties involved when we want to construct the action at higher order in derivatives. 
\begin{itemize}[leftmargin=*]
\item First, whenever we add new operators we want to make sure that we are not strongly breaking the de Sitter dilation symmetry. 
While in the EFT of Inflation this is automatically ensured if we use \sm{\partial_\mu t}, which is invariant under \sm{t\to t+c}, 
things are different if we use \sm{\partial_\mu x^i}, which is not invariant under scale transformations \sm{x^i\to\lambda x^i}.\footnote{Here 
we have in mind that the Greek indices are all contracted in a diffeomorphism-invariant way when we construct operators to put in the Lagrangian 
(e.g.~we form objects like \sm{R^{ii} = R^{\mu\nu}\partial_\mu x^i\partial_\nu x^i}, and so on). 
The resulting operators, then, will not be invariant under dilations unless one divides by an appropriate power of \sm{\sqrt{X}}. 
Equivalently, one can think of \sm{x^i\to\lambda x^i} as an internal transformation, i.e.~a transformation of the three 
scalar fields describing the Lagrangian coordinates of the solid elements.} 
More prosaically, the diffeomorphism-breaking operators should 
be constructed from the unit-norm one-forms \sm{\partial_\mu x^i/\sqrt{X}}, normal to the worldlines \sm{x^i={\rm const.}} 
of the volume elements of the solids. 
\item Second is the problem of tadpole subtraction. Once we include higher-derivative operators, it is not possible to 
reabsorb their background in the leading-order action of \eq{solids-general_action}. Let us consider, for example, the 
equivalent of the extrinsic curvature. Given the worldlines \sm{x^i={\rm const.}}, 
we can construct the unit vector parallel to them as \cite{Dubovsky:2005xd} 
\begin{equation}
\label{eq:solids-unit_vector}
O^\mu = \frac{\vec{e}^{\mu\nu\rho\sigma}\epsilon_{ijk}\partial_\nu x^i\partial_\rho x^j\partial_\sigma x^k}{6\sqrt{\det(g^{mn})}}\,\,,
\end{equation} 
where \sm{\vec{e}_{\mu\nu\rho\sigma}={\sqrt{-g}\,\epsilon_{\mu\nu\rho\sigma}}} (\sm{\vec{e}^{\mu\nu\rho\sigma}={-\epsilon_{\mu\nu\rho\sigma}/\sqrt{-g}}}) 
is the volume form. Given the projector \sm{{\cal H}_{\mu\nu} = g_{\mu\nu} + O_\mu O_\nu}, the tensor 
\begin{equation}
\label{eq:solids-extrinsic_curvature-A}
{\cal K}_{\mu\nu} = {\cal H}_{(\mu}^{\hp{(\mu}\rho}\nabla_\rho O_{\nu)} 
\end{equation}
is then the equivalent of \sm{K_{\mu\nu}}, to which it would reduce if \sm{\vec{e}^{\mu\nu\rho\sigma}O_\nu\nabla_\rho O_\sigma=0}.\footnote{{Again,} we 
refer to \cite{Dubovsky:2005xd,Endlich:2010hf,Gripaios:2014yha} for a more detailed discussion about vorticity, which is outside of the scope of this work.} 
It is now impossible, however, to write its background without breaking time diffeomorphisms. 
One can construct an object starting at linear order in perturbations by taking the combination 
\begin{equation} 
\label{eq:extrinsic_curvature-B} 
\delta {\cal K}_{\mu\nu} = {\cal K}_{\mu\nu} - \frac{\nabla_\rho O^\rho}{3} {\cal H}_{\mu\nu}\,\,. 
\end{equation}
However when we construct, for example, the operator \sm{{\cal K}_{\mu\nu}{\cal K}^{\mu\nu}}, we run into the issue that the operator \sm{(\nabla_\rho O^\rho)^2} 
is not contained in the action of \eq{solids-general_action}. This way of phrasing the problem suggests its solution: it should be enough to add 
a dependence of \sm{F} on a single additional operator, \sm{\nabla_\rho O^\rho}. 

While the problem of tadpole subtraction is immaterial if one wants to study 
graviton interactions only, as we will see in detail in Section~\ref{sec:main_section}, 
it is still worth to keep it in mind as a difference between this symmetry breaking pattern and 
the EFT of Inflation. {We elaborate on this in Appendices~\ref{app:appendix-new} and \ref{app:appendix-A}.} 
\end{itemize}

\subsection{\texorpdfstring{Type-II framid ($\supset$ type-I framid)}{Type-II framid (\textbackslash supset type-I framid)}} 
\label{subsec:framids}

\noindent Let us first focus on the case of type-II framids. As it is easy to imagine, the fact that we do not have a clock tells us 
that we cannot write an effective theory for perturbations around a generic FLRW spacetime. 

Thus, if we wanted to use this symmetry breaking 
pattern as a legit inflationary model, we should add new degrees of freedom to exit inflation (the simplest thing to do is to add a clock, which 
would result in the symmetry breaking pattern of type-II superfluids). 
Let us nevertheless press forward and see how one can obtain a pure de Sitter solution, since this allows us to introduce the geometric objects that will be used in 
Section~\ref{sec:main_section}. 

The quickest way to show this is by looking at the different operators one can write down, order-by-order in derivatives. 
First, given a dreibein \sm{e^\mu_I}, we can immediately obtain the unit vector orthogonal to \sm{e^\mu_I} via the volume form as 
\begin{equation}
\label{eq:framids-1}
e^\mu_0 = -\frac{g^{\mu\nu}}{6}\epsilon_{IJK}\vec{e}_{\nu\alpha\beta\gamma}e^\alpha_Ie^\beta_Je^\gamma_K\,\,. 
\end{equation} 
Importantly, this unit vector is not only orthogonal to \sm{e^\mu_I}, but one can show that if we define the tetrad as \sm{e^\mu_A=\{e^\mu_0,e^\mu_I\}}, 
and transform \sm{e^\mu_I} as \sm{e'^\mu_I = \Lambda_I^{\hp{I}B}e^\mu_B} (where \sm{\Lambda_A^{\hp{A}B}} is a local Lorentz transformation), we have that 
\begin{equation}
\label{eq:framids-2}
e'^\mu_0 = -\frac{g^{\mu\nu}}{6}\epsilon_{IJK}\vec{e}_{\nu\alpha\beta\gamma}e'^\alpha_Ie'^\beta_Je'^\gamma_K = \Lambda^A_0e^\mu_A\,\,. 
\end{equation} 
What operators can we construct from the tetrad \sm{e^\mu_A}, then? 
\begin{itemize}[leftmargin=*]
\item At zeroth order in derivatives, the only non-gauge-invariant quantity we can use is \sm{e^\mu_A} itself. Since greek indices must be contracted, the 
only object we can use is \sm{g_{\mu\nu}e^\mu_Ae^\nu_B = \eta_{AB}}, 
where tetrad indices must be contracted in an \sm{SO(3)}-invariant way. These only modify the cosmological constant. 
\item At first order in derivatives we have four additional building blocks. These are the tensors 
\begin{equation}
\label{eq:framids-3}
e^\nu_0\nabla_\nu e^\mu_0\,\,, \quad e^\nu_0\nabla_\nu e^\mu_J\,\,, \quad e^\nu_I\nabla_\nu e^\mu_0\,\,, \quad e^\nu_I\nabla_\nu e^\mu_J\,\,. 
\end{equation} 
It is straightforward to see that around a de Sitter spacetime we can subtract the background 
of these tensors without introducing new degrees of freedom. More precisely, defining the dual one-forms 
\begin{equation}
\label{eq:framids-4}
\omega_\mu^A=\eta_{AB}g_{\mu\nu}e^\nu_B\,\,,
\end{equation} 
we can isolate the fluctuations as 
\begin{subequations}
\label{eq:framids-5}
\begin{align}
\delta(e^\nu_I\nabla_\nu e^\mu_0) &= e^\nu_I\nabla_\nu e^\mu_0 - H e^\mu_I\,\,, \label{eq:framids-5-1} \\
\delta(e^\nu_I\nabla_\nu e^\mu_J) &= e^\nu_I\nabla_\nu e^\mu_J - H\delta_{IJ}e^\mu_0 
\label{eq:framids-5-2} 
\end{align}
\end{subequations} 
and 
\begin{equation} 
\label{eq:framids-6} 
\delta(\nabla_\nu e^\mu_0) = \nabla_\nu e^\mu_0 - H(\delta_\nu^{\hp{\nu}} - \omega_\nu^0 e^\mu_0)\,\,. 
\end{equation} 
These imply that \sm{e^\nu_0\nabla_\nu e^\mu_0} vanishes for a de Sitter metric, as does \sm{e^\nu_0\nabla_\nu e^\mu_J}. 
\item At higher order in derivatives, via the product rule we can always rewrite the background of non-gauge-invariant 
tensors in terms of \eqsII{framids-5}{framids-6}. 
\end{itemize} 
One can also check how these objects transform under the de Sitter dilation symmetry. It is easy to see that, in de Sitter spacetime, 
the Lie derivative of \sm{e^\mu_I} along the vector field \sm{\delta^\mu_0 - Hx^i\delta^\mu_i} is equal to zero. 

What about type-I framids? Given \eqsII{framids-1}{framids-2}, we see that this symmetry breaking pattern is a subset of the type-II framid: 
for this reason we will not discuss them further. It is interesting to notice that if we require a symmetry under the global transformation 
\begin{equation}
\label{eq:framids-7} 
e^\mu_I\to M_I^{\hp{I}J}e^\mu_J\,\,, 
\end{equation} 
with \sm{M_I^{\hp{I}J}} a three-dimensional matrix having unit determinant, it seems we can single out \sm{e^\mu_0} 
as the only vector from which to construct non-gauge-invariant operators. However this works only at leading (zeroth) order 
in derivatives. Given that finding such symmetry is not necessary for the purposes of this paper, we leave this question to future work.

\subsection{A word on the Goldstone modes} 
\label{subsec:a_word_on_Goldstone_modes}

\noindent Before proceeding, we find it useful to quickly summarize how the Stueckelberg trick 
to reintroduce the Goldstone modes works for these different symmetry breaking patterns. 
\begin{itemize}[leftmargin=*]
\item In the superfluid case we are breaking time diffeomorphisms. The Stueckelberg trick, then, 
simply amounts to performing a time diffeomorphism \sm{t\to t+\pi}, as explained in detail in \cite{Cheung:2007st}. 
After the Stueckelberg trick, the ``flat-gauge metric'' can be written in the usual ADM decomposition, with a spatial 
part that only contains the graviton (i.e.~no vector or scalar modes). 
\item In the case of a solid we are breaking spatial diffeomorphisms. The Stueckelberg trick is a spatial diffeomorphism \sm{x^i\to x^i+\pi^i}, where 
\sm{\pi^i} can be decomposed in a scalar and a vector part. Again, the spatial metric only contains the graviton after performing the Stueckelberg trick. 
\item In the case of framids the Stueckelberg trick consists in doing a local Lorentz transformation. As we have already discussed 
in Section~\ref{subsec:sixfold}, the parameters of this transformation are the rapidity and the three Euler angles, i.e.~we have six degrees of 
freedom (two scalars and two vectors). Also in this case the metric after the Stueckelberg trick can be decomposed in ADM variables, with 
only the graviton present in \sm{g_{ij}} (see e.g.~\cite{Delacretaz:2015edn} for the case of type-I framids). 
\end{itemize} 
{In this work we consider the minimal implementations of the different symmetry breaking patterns. 
Consequently we never have additional spin-\sm{2} degrees of freedom besides the graviton.} 
{There are scenarios where these additional degrees of freedom are present: one example is 
Gaugid Inflation \cite{Piazza:2017bsd}, which belongs to the symmetry breaking pattern 
of a solid. However, it is important to emphasize that in this model the additional spin-\sm{2} mode, \sm{E_{ij}}, 
behaves differently from the graviton: its quadratic action is not of the form 
\sm{{-\int}\dif^4x\,a^3\,g^{\mu\nu}\partial_\mu E_{ij}\partial_\nu E_{ij}}. 
For this reason we have not discussed this scenario in more detail. Regarding this, it is also worth 
to emphasize that the construction of \cite{Bordin:2018pca}, in which the authors use the breaking 
of de Sitter boosts to add light spinning particles to the EFT of Inflation (see e.g.~their Section 5), 
could be extended to the different scenarios discussed in this paper.}

\section{Main results} 
\label{sec:main_section}

\noindent In this section we collect the main results of this paper. First, we show that the 
solid is the symmetry breaking that allows the most freedom for graviton interactions (Section \ref{subsec:most_general_graviton_interactions}). 
Then, we study some phenomenological consequences of these interactions (Section \ref{subsec:a_bit_of_phenomenology}).

\subsection{Most general graviton interactions} 
\label{subsec:most_general_graviton_interactions}

\noindent The best way to see what is the symmetry breaking pattern that allows the most 
freedom is working at the level of linear fluctuations. 
That is, we work directly with the graviton field \sm{\gamma_{ij}}, and check what are the 
``minimal'' building blocks in the different symmetry breaking patterns. 

Since we only focus on the graviton, we can take the line element to be 
\begin{equation}
\label{eq:most_general_graviton_interactions-metric}
\dif s^2=-\dif t^2 + a^2(\eu^{\gamma})_{ij}\dif x^i\dif x^j\,\,, 
\end{equation} 
that is we set the lapse and shift variables to zero, together with the scalar and vector modes that are introduced by making broken diffeomorphisms or 
broken local Lorentz transformations (see Section~\ref{subsec:a_word_on_Goldstone_modes} for details).\footnote{When looking at interactions higher 
than cubic one also needs to include the constraints (which 
otherwise need only to be solved at linear order \cite{Maldacena:2002vr}, and consequently do not contain \sm{\gamma_{ij}}). These terms 
do not change the conclusions of this paper, so we leave their detailed study to future work.} 
From this, we can write the tetrad as 
\begin{equation}
\label{eq:most_general_graviton_interactions-tetrad}
e^\mu_0 = \delta^\mu_0\,\,, \quad e^0_I = 0\,\,, \quad e^i_I = a^{-1}(\eu^{-\frac{\gamma}{2}})_{iI}\,\,. 
\end{equation} 

\begin{description}[leftmargin=0pt,labelwidth=0pt] 
\item[Type-I superfluids] Let us start from looking at the building blocks for the EFT of Inflation. 
At linear order, \sm{\gamma_{ij}} can only appear with derivatives: time derivatives enter 
as \sm{\delta\!K_i^{\hp{i}j} = \dot{\gamma}_{ij}/2}, while spatial derivatives \sm{a^{-2}\partial_k\partial_l\gamma_{ij}} enter via 
\sm{{^{(3)}}\!R^{ij}_{\hp{ij}kl}}, the Riemann curvature tensor on 
constant-time hypersurfaces. More precisely, \sm{a^{-2}\partial_k\partial_l\gamma_{ij}} can only appear in the combination 
\begin{equation}
\label{eq:most_general_graviton_interactions-curvature_tensor}
{^{(3)}}\!R^{ij}_{\hp{ij}kl} = a^{-2}\big(\partial_k\partial_{[i}\gamma_{j]l} - \partial_l\partial_{[i}\gamma_{j]k}\big)\,\,. 
\end{equation} 
It is then possible to take any number of spatial and time derivatives of these building blocks by 
using the covariant derivative \sm{D_\mu} on the \sm{t={\rm const.}} hypersurfaces and \sm{n^\mu\nabla_\mu}, respectively. 
\item[Type-II framids] We then move to the type-II framid. First, we see that using \sm{e^\mu_0} allows us to reproduce all the geometric objects of the 
EFT of Inflation. Then, given that at lowest order in derivatives we cannot write any non-gauge-invariant object -- we 
always end up with a cosmological constant -- it is still impossible to write \sm{\gamma_{ij}} without derivatives. Unlike the case of broken time diffeomorphisms, however, it is now possible to use 
\sm{a^{-1}\partial_k\gamma_{ij}} as a building block. More precisely, we can use the combination 
\begin{equation}
\label{eq:most_general_graviton_interactions-framid_christoffel} 
{-\omega_\mu^K} \delta(e^\nu_J\nabla_\nu e^\mu_I) = {-\omega_\mu^K} e^\nu_J\nabla_\nu e^\mu_I = {a^{-1}}\partial_{[K}\gamma_{I]J}\,\,. 
\end{equation} 
Here we notice the overall factor of \sm{a^{-1}}, that we expect from the discussion on 
invariance under de Sitter dilations of Section~\ref{subsec:framids}. 
\item[Solids] When we break spatial diffeomorphisms we can still reproduce all the operators of the EFT of Inflation. 
This is achieved by using the 
vector \sm{O^\mu} defined in \eq{solids-unit_vector}. Let us go back to the problem of the cancellation of tadpoles. We know that in the EFT of 
Inflation the background of all operators can be constructed from the normal vector to the constant-time hypersurfaces 
(whose role here is played by \sm{O^\mu}), and the Hubble rate with its time derivatives. We have seen, e.g.~in \eq{extrinsic_curvature-B}, 
that the four-divergence of \sm{O^\mu} can play the role of the Hubble rate. We can then use \sm{O^\mu\nabla_\mu} to take time derivatives of this proxy for \sm{H}. 
There is no loss of generality in doing this if we can show that these objects do not contain the graviton. Luckily, this is easy to prove. From the 
definition of \eq{solids-unit_vector}, we see that \sm{O^\mu} does not contain \sm{\gamma_{ij}} since neither \sm{\vec{e}^{\mu\nu\rho\sigma}} 
nor \sm{\sqrt{\det(g^{mn})}} do. Using the formula 
\begin{equation}
\label{eq:most_general_graviton_interactions-divergence_formula}
\nabla_\mu O^\mu = \frac{1}{\sqrt{{-g}}}\,\partial_\mu(\sqrt{{-g}}\,O^\mu) 
\end{equation} 
we see that its four-divergence also does not contain the graviton. 
Then, given that to take its time derivatives we are only acting with \sm{O^\mu\nabla_\mu} on a scalar, 
we can replace \sm{\nabla_\mu} with \sm{\partial_\mu} (which also does not contain \sm{\gamma_{ij}}). 

What about spatial derivatives? It is straightforward to see that we can use directly \sm{{a^{-1}}\partial_{k}\gamma_{ij}} (where, 
unlike \eq{most_general_graviton_interactions-framid_christoffel}, we do not need to take any antisymmetrization over \sm{k} and \sm{j}) by considering 
\begin{equation}
\label{eq:most_general_graviton_interactions-solid_christoffel-1}
\Gamma^{kij}\equiv-\frac{g^{\mu k}g^{\nu (i}\nabla_\mu\nabla_\nu x^{j)}}{(X/3)^{{3}/{2}}}\,\,, 
\end{equation}
a quantity that starts linear in perturbations. As discussed in Section~\ref{subsec:solids}, we divide by the appropriate power of \sm{X} to ensure invariance 
under spatial dilations. This in turn ensures that we reproduce the expected power of the scale factor when we have one spatial derivative acting on \sm{\gamma_{ij}}, as 
in \eq{most_general_graviton_interactions-framid_christoffel}. Indeed, we find 
\begin{equation}
\label{eq:most_general_graviton_interactions-solid_christoffel-2}
\frac{1}{2a}\partial_{k}\gamma_{ij} = \Gamma^{kij}\,\,. 
\end{equation}

Finally, there is one more additional freedom with respect to the case of type-II framids. While there we can only contract spatial indices 
via the Kronecker delta, here we can use \sm{\gamma_{ij}} itself through 
\begin{equation}
\label{eq:most_general_graviton_interactions-solid-more_freedom}
\Gamma^{ij}\equiv\delta_{ij} - \frac{3g^{ij}}{X}\,\,, 
\end{equation} 
which is equal to \sm{\gamma_{ij}} at linear order in perturbations. 
\end{description}

\subsection{A bit of phenomenology} 
\label{subsec:a_bit_of_phenomenology}

\noindent Let us then discuss the phenomenology of the solid symmetry breaking pattern. Given the large amount of freedom we have, 
characterizing all the signatures is a complicated task. For this reason, we think it is more useful to compare with the EFT of Inflation. 

While we have shown that the symmetry breaking pattern of a solid captures the most general graviton interactions 
at \emph{all} orders in perturbations and derivatives, when discussing the phenomenology it is useful to work in a derivative expansion. 
We stop at second order in this expansion. Moreover, we focus on operators that start from cubic order in perturbations at most, i.e.~we 
look at the tree-level graviton bispectrum. 

At lowest (zeroth) order in derivatives, we have a small mass of order 
\sm{\varepsilon H^2} for the graviton, and a cubic vertex \sm{\gamma_{ij}\gamma_{jk}\gamma_{ki}} that instead 
is not slow-roll-suppressed (see \cite{Endlich:2012pz,Endlich:2013jia}, for example: we review this in 
Appendix~\ref{app:appendix-B}). The latter has been shown in \cite{Endlich:2013jia} 
to give a graviton bispectrum that peaks in the squeezed limit, and that does not asymptote a constant at late times 
even at leading order in \sm{\varepsilon} (at this order we can use de Sitter modes to compute the in-in integral: given that the interaction is purely local, it is 
easy to see that there are divergences of the form \sm{\ln({-k\eta})}, where \sm{\eta} is the conformal time \sm{-1/aH}). 

Let us go to second order in derivatives (i.e.~consider the new operators discussed in Section~\ref{subsec:most_general_graviton_interactions}). 
At quadratic order in perturbations, we can look at the different ways in which the speed of sound of the graviton can be modified. {First, using 
$O^\mu$ we can construct the operators of the EFT of Inflation} like the Ricci scalar on the constant-time hypersurfaces \sm{\tr} 
and the square of the extrinsic curvature \sm{\delta {\cal K}_{\mu\nu}\delta {\cal K}^{\mu\nu}}{. We have already discussed 
below \eq{solids-extrinsic_curvature-A} how ${\cal K}_{\mu\nu}$ reduces to the extrinsic curvature when only tensor perturbations are considered. 
In a similar way it is possible to construct an object that reduces to \sm{\tr}: we discuss this in Appendix~\ref{app:appendix-C}.} 
{Besides these,} there are many new operators that reduce to \sm{\dot{\gamma}_{ij}^2} and \sm{a^{-2}(\partial_k\gamma_{ij})^2} 
at quadratic order in \sm{\gamma_{ij}}, but differ from the EFT of Inflation ones at cubic order (and higher). Two simple examples are 
\begin{equation}
\label{eq:a_bit_of_phenomenology-simple_speed_of_sound}
(O^\mu\nabla_\mu\Gamma^{ij})(O^\nu\nabla_\nu\Gamma^{ij})\,\,,\quad\Gamma^{kij}\Gamma^{kij}\,\,. 
\end{equation} 
An operator that instead is essentially equivalent to the three-Ricci scalar is the one studied in \cite{Ricciardone:2016lym}, 
i.e.~\sm{G^{ii}}, where \sm{G_{\mu\nu}} is the four-dimensional Einstein tensor (indeed, Ref.~\cite{Ricciardone:2016lym} 
shows that the resulting graviton bispectrum has the same shape as that from Einstein gravity). 

A phenomenologically interesting observation comes when we turn from the quadratic action to the interactions. One of the main points of \cite{Creminelli:2014wna} 
is that by appropriately redefining the lightcone and consequently putting the tensor sound speed to \sm{1}, one also removes completely 
all graviton self-interactions beyond the ones from Einstein gravity. This relies on the fact that there are only two operators to remove, 
\sm{\tr} and \sm{\delta {\cal K}_{\mu\nu}\delta {\cal K}^{\mu\nu}}, and two free coefficients in the lightcone redefinition 
(those of a conformal and disformal transformation of the metric, see also \cite{Bordin:2017hal} for details). 
Now the situation is different simply because we have more operators to play with, but the same number of free coefficients in the lightcone 
redefinition.\footnote{Indeed, at this order in derivatives and perturbations the only objects we can use 
are the vector \sm{O^\mu} and the normalized one-forms \sm{\partial_\mu x^i/\sqrt{X}}. Because we need to contract latin 
indices in an \sm{SO(3)}-invariant way, 
we can only use the combination \sm{\partial_\mu x^i\partial_\nu x^i/X}, which is essentially equivalent to \sm{g_{\mu\nu} + O_\mu O_\nu}.} 
This tells us that, while there is no loss of generality in putting the tensor 
speed of sound to \sm{1}, doing so does \emph{not} remove also the interaction terms. 

Another important phenomenological difference with the EFT of Inflation is the following. 
As we have seen, all interactions in that case come from operators starting quadratic in \sm{\gamma_{ij}}: there 
are no operators starting at cubic order in \sm{\gamma_{ij}} at second order in derivatives. 
If we break spatial diffeomorphisms the situation changes. For example, we can combine 
\eq{most_general_graviton_interactions-solid-more_freedom} with \eq{a_bit_of_phenomenology-simple_speed_of_sound} to construct 
\begin{equation}
\label{eq:a_bit_of_phenomenology-purely_cubic_interactions}
\text{$\gamma_{jl}\dot{\gamma}_{ij}\dot{\gamma}_{il}$ \quad and \quad $a^{-2}\gamma_{kl}\partial_k\gamma_{ij}\partial_l\gamma_{ij}\,\,,$} 
\end{equation}
where we notice that the second structure appears directly in the Einstein-Hilbert action 
(it is contained in \sm{\tr} together with \sm{a^{-2}\gamma_{jl}\partial_k\gamma_{ij}\partial_k\gamma_{il}}), while the first one does not. 
Most importantly, both structures contain an undifferentiated tensor mode. 
This confirms that, if spatial diffeomorphisms are broken, the squeezed limit ${\cal O}(1/q^3)$ of the three-point function of the graviton is not 
controlled only by coefficients entering in the graviton power spectrum. This is already apparent at the zeroth order in derivatives 
(for example, we can see this by comparing \eq{appendix-B-3} with \eq{appendix-B-4} in Appendix~\ref{app:appendix-B}): here we arrive 
at the same conclusion at second order. This is a fast way to see that the consistency relations are not respected if we break spatial 
diffeomorphisms. We refer to \cite{Endlich:2013jia,Bordin:2017ozj,Pajer:2019jhb} 
for a discussion of soft limits involving scalar and vector modes in Solid Inflation. 

It is straightforward to compute the graviton three-point function from the operators of 
\eq{a_bit_of_phenomenology-purely_cubic_interactions}. Using the standard Fourier decomposition for the graviton,\footnote{
This is
\begin{equation}
\label{eq:bispectra-A}
\gamma_{ij}(t,\vec{x}) = \int\frac{\dif^3k}{(2\pi)^3}\sum_s\epsilon^s_{ij}(\vec{k})\gamma^{s}_{\vec{k}}(t)\eu^{\iu\vec{k}\cdot\vec{x}}\,\,, 
\end{equation} 
where the traceless polarization tensors \sm{\epsilon^s_{ij}} satisfy \sm{k^i\epsilon^s_{ij}(\vec{k})=0}.} 
together with the three elementary symmetric polynomials 
\begin{subequations}
\label{eq:bispectra-B}
\begin{alignat}{3}
k_T &= e_1 &&= k_1+k_2+k_3\,\,, \label{eq:bispectra-B-1} \\ 
{} &\,\,\,\,\,\,\,\, e_2 &&= k_1k_2+k_2k_3+k_3k_1\,\,, \label{eq:bispectra-B-2} \\ 
{} &\,\,\,\,\,\,\,\, e_3 &&= k_1k_2k_3\,\,, \label{eq:bispectra-B-3} 
\end{alignat}
\end{subequations} 
we find 
\begin{subequations}
\label{eq:bispectra-C}
\begin{align}
\braket{\gamma^{s_1}_{\vec{k}_1}\gamma^{s_2}_{\vec{k}_2}\gamma^{s_3}_{\vec{k}_3}}'_{\gamma\dot{\gamma}\dot{\gamma}} &= 
\frac{H^4M^2}{2\mpl^6}\,\frac{k_Te^2_2+e_2e_3-2k_T^2e_3}{e_3^3k_T^2}\, 
\epsilon^{s_1}_{ij}({-\vec{k}_1})\epsilon^{s_2}_{jk}({-\vec{k}_2})\epsilon^{s_3}_{ki}({-\vec{k}_3})\,\,, \label{eq:bispectra-C-1} \\
\braket{\gamma^{s_1}_{\vec{k}_1}\gamma^{s_2}_{\vec{k}_2}\gamma^{s_3}_{\vec{k}_3}}'_{\gamma\partial{\gamma}\partial{\gamma}} &= 
\frac{H^4M^2}{2\mpl^6}\,\frac{k_Te_2+e_3-k_T^3}{e_3^3k_T^2}\, 
\epsilon^{s_1}_{il}({-\vec{k}_1})\epsilon^{s_2}_{jm}({-\vec{k}_2})\epsilon^{s_3}_{kn}({-\vec{k}_3})\, 
T_{ijk}^{lmn}(\vec{k}_1,\vec{k}_2,\vec{k}_3)\,\,, \label{eq:bispectra-C-2}
\end{align}
\end{subequations} 
with 
\begin{equation}
\label{eq:bispectra-D}
T_{ijk}^{lmn}(\vec{k}_1,\vec{k}_2,\vec{k}_3) = 
k_2^ik_3^l\delta_{jk}\delta_{mn} + k_3^jk_1^m\delta_{ki}\delta_{nl} + k_1^kk_2^n\delta_{ij}\delta_{lm}\,\,. 
\end{equation}
Here \sm{M^2} is a mass scale that makes the operators have mass dimension \sm{4}. We have used de Sitter 
mode functions to compute these bispectra: given that the mass of the graviton is slow-roll-suppressed, this is 
correct at leading order in \sm{\varepsilon}.

A final observation that is worth making is the following. If we break parity, in the EFT of Inflation it is only 
possible to modify the quadratic action of the graviton starting from third order in derivatives, via operators that reduce to 
\begin{equation}
\label{eq:a_bit_of_phenomenology-parity_violation-A}
\text{$a^{-1}\epsilon_{ijk}\partial_i\dot{\gamma}_{jl}\dot{\gamma}_{lk}$ \quad and \quad 
$a^{-3}\epsilon_{ijk}\partial_i\partial_l\gamma_{jm}\partial_l\gamma_{mk}$} 
\end{equation} 
at quadratic order in \sm{\gamma_{ij}}. Now it is possible to write operators both at first and second order in derivatives, via 
\begin{subequations}
\label{eq:a_bit_of_phenomenology-parity_violation-B}
\begin{align}
\epsilon_{ijk} \Gamma^{il}\Gamma^{jlk} &= a^{-1}\epsilon_{ijk}\gamma_{il}\partial_j\gamma_{lk}\,\,, 
\label{eq:a_bit_of_phenomenology-parity_violation-B-1} \\
\epsilon_{ijk} \Gamma^{jlk}O^\mu\nabla_\mu\Gamma^{il} &= a^{-1}\epsilon_{ijk}\dot{\gamma}_{il}\partial_j\gamma_{lk}\,\,. 
\label{eq:a_bit_of_phenomenology-parity_violation-B-2}
\end{align}
\end{subequations} 
These two operators split the graviton helicities. However, one can easily see that they lead to 
serious instabilities on subhorizon scales. Given the circularly-polarized polarization tensors 
\sm{\epsilon^\pm_{ij}(\vec{k})}, and using the property \sm{\iu k^l\epsilon_{jlm}\epsilon^s_{im}(\vec{k}) 
= \lambda_s k\epsilon^s_{ij}(\vec{k})}, \sm{\lambda_\pm = \pm 1}, we see that the contribution of these two operators to the dispersion 
relation of the two \mbox{helicities is} 
\begin{subequations}
\label{eq:a_bit_of_phenomenology-parity_violation-C}
\begin{align}
\omega^2 &= k^2 + \lambda_sMk\,\,, \label{eq:a_bit_of_phenomenology-parity_violation-C-1} \\
\omega^2 &= k^2 + \iu c\lambda_s\omega k\,\,. \label{eq:a_bit_of_phenomenology-parity_violation-C-2}
\end{align}
\end{subequations}
Here \sm{M} and \sm{c} (of mass dimension \sm{1} and \sm{0} respectively) are the coefficients controlling the operators of 
\eq{a_bit_of_phenomenology-parity_violation-B}, we have neglected the mass of the graviton since it is of order 
\sm{\varepsilon H^2}, and we assume that \sm{M\gg H}. 
We see that the second operator essentially gives a complex speed of sound to the graviton (one can see this in the limit \sm{c\ll 1}), 
while the first operator makes the frequency of one of the two helicities purely imaginary for \sm{k\ll M}. 
In summary, these operators lead to instabilities on short scales, and it is therefore unclear how to make sense 
of the Bunch-Davies vacuum if we turn them on. 
This is also why we do not study parity-breaking operators at higher order in perturbations: 
it does not make much sense to do so while putting to zero the operators of \eqsI{a_bit_of_phenomenology-parity_violation-B}, 
since the cubic and higher-order operators could generate these via loop effects.\footnote{Indeed, 
there is no symmetry besides parity that is recovered if we put \sm{M} or \sm{c} to zero.} 
We leave a more detailed investigation of this issue for a future work.

\section{Conclusions} 
\label{sec:conclusions}

\noindent In this paper we have studied graviton non-Gaussianities in models of the early universe 
in which de Sitter boosts are spontaneously broken, showing that the case where the epoch of de Sitter 
expansion is driven by a solid is the one with the more freedom for graviton self-interactions. 

This freedom makes systematically studying all the possible correlation functions very complicated, and we expect 
that a full characterization of non-Gaussianities involving also the Goldstone modes (a scalar and a vector for a solid) 
to be even more difficult. For this reason it would be interesting to extend the approach of the ``Boostless Bootstrap'' 
recently put forward in \cite{Pajer:2020wxk} to this symmetry breaking pattern (recently, Ref.~\cite{Cabass:2021fnw} has made progress in this direction). 
The well-known fact that cosmological perturbations are not conserved in Solid Inflation, even at leading order in the slow-roll 
expansion (for the graviton this is exemplified by the unsuppressed cubic interaction 
\sm{\propto\gamma_{ij}\gamma_{jk}\gamma_{ki}}, and see e.g.~\cite{Bartolo:2013msa,Akhshik:2014gja,Bartolo:2014xfa,
Dimastrogiovanni:2014ina,Akhshik:2014bla,Abolhasani:2015cve,Bordin:2016ruc,Meszaros:2019tfd} 
for studies of mixed three-point functions involving curvature perturbations), makes it difficult to set up a theory 
of correlators on the boundary of de Sitter: we leave investigations of such issues to future work. 

Finally, we emphasize that in this work we have focused on \emph{isotropic} solids, i.e.~where we have invariance under 
the full \sm{SO(3)} group. Refs.~\cite{Kang:2015uha,Kang:2018bqc,Nicolis:2020rqz} study what happens if this assumption 
is relaxed to invariance under a discrete subgroup of rotations. It would be interesting to study how our results generalize 
to their setup.

\section*{Acknowledgements}

\noindent It is a pleasure to thank Misha Ivanov, Enrico Pajer, Gianmassimo Tasinato, Matias Zaldarriaga, 
and especially Mehrdad Mirbabayi and Sadra Jazayeri for useful discussions. I also thank Mehrdad Mirbabayi 
and Enrico Pajer for very useful comments on the draft. I acknowledge support from the Institute for Advanced Study.

\appendix

\section{{About tadpole subtraction}}
\label{app:appendix-new}

\noindent As emphasized in the main text, and made clear in the original paper \cite{Cheung:2007st}, a key ingredient of the EFT of Inflation is 
the tadpole cancellation. This is what allows to write an effective theory for metric fluctuations, which are the quantities we constrain with 
late-time cosmological observations, dependent on free EFT coefficients \emph{and} 
the expansion history during inflation. Isolating all the non-gauge-invariant operators, as done e.g.~in \cite{Lin:2015cqa} 
(and as we have also done in Sections~\ref{subsec:solids} and \ref{subsec:most_general_graviton_interactions}), is not enough. 

Can we achieve this in the case of broken spatial diffeomorphisms? We have seen in Section~\ref{subsec:solids} that 
we can use \sm{\theta\equiv\nabla_\mu O^\mu} to subtract the background of geometric objects constructed from \sm{O^\mu}. 
In this appendix we want to show \emph{why} one has to indeed introduce this new operator beyond $X$, $Y$ and $Z$, by making 
a comparison to the construction of the EFT of Inflation. 

Let us first recall what happens in the case of broken time diffeomorphisms. The chief example is that 
of the extrinsic curvature. Let us consider the operator $K_{\mu\nu}K^{\mu\nu}$ and define $\delta\!K_{\mu\nu}$ as 
$K_{\mu\nu} - Hh_{\mu\nu}$. We can then rewrite $K_{\mu\nu}K^{\mu\nu}$ as 
\begin{equation}
\label{eq:appendix-new-A-1}
K_{\mu\nu}K^{\mu\nu} = \delta\!K_{\mu\nu}\delta\!K^{\mu\nu} + 2Hh_{\mu\nu}\delta\!K^{\mu\nu} + 3 H^2\,\,. 
\end{equation} 
The first term starts at quadratic order in perturbations, while the last term is reabsorbed in the 
``cosmological constant'' in \eq{EFTI-general_action}. We rewrite the middle term as 
\begin{equation}
\label{eq:appendix-new-A-2}
{-6H^2} + 2Hh_{\mu\nu}K^{\mu\nu}\,\,. 
\end{equation} 
The first term here is again reabsorbed by \eq{EFTI-general_action}. What about the second term? One can integrate it by parts to arrive at 
\begin{equation}
\label{eq:appendix-new-A-3}
{-2n^\mu\nabla_\mu H} = {-\frac{2\dot{H}}{N}}\,\,, 
\end{equation} 
where we recall that $g^{00} = {-1/N^2}$. The key point, now, is that the right-hand side is already contained in action at zeroth-order in derivatives, 
i.e.~the slow-roll action plus all the operators of the form \sm{(g^{00}+1)^n}. This example shows how in the case of 
broken time diffeomorphisms we do not need new operators beyond the zeroth-derivative ones if we want to account for all tadpoles. 

We can then move to the case of broken spatial diffeomorphisms. Given an expansion history $H(a)$, we can find 
a function $f(X)$ equal to $H(a)$ on the background, where $X=3/a^2$. Then, mirroring what we did for the case of 
time spatial diffeomorphisms, we consider the operator \sm{{\cal K}_{\mu\nu}{\cal K}^{\mu\nu}}, where 
${\cal K}_{\mu\nu}$ is given by \eq{solids-extrinsic_curvature-A}. If we define 
\begin{equation}
\label{eq:appendix-new-B-1} 
\delta{\cal K}_{\mu\nu} = {\cal K}_{\mu\nu} - f(X){\cal H}_{\mu\nu}\,\,, 
\end{equation} 
which vanishes on the background, we find 
\begin{equation}
\label{eq:appendix-new-B-2}
{\cal K}_{\mu\nu}{\cal K}^{\mu\nu} = \delta{\cal K}_{\mu\nu}\delta{\cal K}^{\mu\nu} + 2 f(X){\cal H}_{\mu\nu}\delta{\cal K}^{\mu\nu} + 3f^2(X)\,\,. 
\end{equation} 
The first piece starts at quadratic order in perturbations, while the last piece is a function of $X$ only, so it already included in 
the action at zeroth order in derivatives -- the Solid Inflation action of \eq{solids-general_action}. We rewrite the middle piece as 
\begin{equation}
\label{eq:appendix-new-B-3}
{-6 f^2(X)} + 2f(X){\cal H}_{\mu\nu}{\cal K}^{\mu\nu}\,\,. 
\end{equation}
The first term is reabsorbed by \eq{solids-general_action}, and we can integrate by parts the second term. However, we now end up with 
\begin{equation}
\label{eq:appendix-new-B-4}
O^\mu\nabla_\mu f(X)\,\,. 
\end{equation} 
Unlike what happens for broken time diffeomorphisms, this operator is \emph{not} included in the action of \eq{solids-general_action}. 
This shows how, if we want to subtract the tadpoles from operators that involve derivatives acting on the unitary-gauge metric, 
we cannot only consider functions of $X$ and \eq{solids-general_action}. Then, it is a matter of choosing the simplest operator to add to 
\eq{solids-general_action} to account for this. In this work we have chosen $\theta$ 
because it does not contain tensor perturbations, as discussed in Section~\ref{subsec:most_general_graviton_interactions}.

\section{\texorpdfstring{An ``EFT of Inflation'' for solids}{An "EFT of Inflation" for solids}}
\label{app:appendix-A}

{Having confirmed that new operators are needed to subtract tadpoles, we are in the position to construct an EFT for broken spatial diffeomorphisms. 
First, the discussion of the previous appendix holds only up to second order in derivatives. Here we are more precise and characterize the operators necessary 
to subtract tadpoles at all orders in derivatives. Then we also show how to solve the background equations of motion when these operators are included 
(for the case of zeroth and second order in derivatives only).} 
\begin{description}[leftmargin=0pt,labelwidth=0pt]
\item[Zeroth order in derivatives] {Let us start from the simplest case.} At this order we have Solid Inflation. Using the fact that 
the derivative of \sm{Y} and \sm{Z} with respect to \sm{g^{\mu\nu}} vanishes for an FLRW metric, there is no loss of generality if 
we rewrite the non-Einstein-Hilbert part of \eq{solids-general_action} as 
\begin{equation}
\label{eq:appendix-A-1}
\int\dif^4x\,\sqrt{-g}\,\big\{{\cal L}_0(X) + M^4(X,\delta Y,\delta Z)\big\}\,\,, 
\end{equation} 
where we have defined \sm{\delta Y \equiv Y-1/3}, \sm{\delta Z \equiv Z-1/9}. 
The form of \sm{{\cal L}_0} is fixed at any given order in slow-roll 
parameters by solving the background Einstein equations. They take the form 
\begin{subequations}
\label{eq:appendix-A-2}
\begin{align}
3\mpl^2H^2 &= {-{\cal L}_0}\,\,, \label{eq:appendix-A-2-1} \\
{-\frac{\dot{H}}{H^2}} &= \frac{\dif\ln{\cal L}_0}{\dif\ln X}\,\,. \label{eq:appendix-A-2-2}
\end{align}
\end{subequations}
Then, we have in mind that the function \sm{M^4} is expanded in powers of 
\sm{\delta Y} and \sm{\delta Z}, with coefficients depending on \sm{X}. This dependence should be mild if we want the de Sitter 
dilation symmetry to be softly broken. This expansion is similar to that of the EFT of Inflation in powers of \sm{g^{00}+1}, 
with EFT coefficients that depend only mildly on time. From now on we do not worry about \sm{M^4} or any other terms in the 
Lagrangian that start at second order in perturbations. 
\item[Up to second order in derivatives] Let us see how at this order we can get away with just adding a dependence of \sm{{\cal L}_0} on \sm{\theta}. 
The only operators that are not accounted are those of the form \sm{O^\mu\nabla_\mu\theta} times a function of \sm{X}. We can integrate these by parts, and then 
use the fact that we can rewrite \sm{{\theta}} {times} \sm{O^\mu\nabla_\mu X} as a function of \sm{X} times \sm{\theta^2} plus terms starting quadratic in perturbations. 
What do the equations of motion look like? Using the relation (valid for any vector \sm{V^\mu})
\begin{equation}
\label{eq:appendix-A-3}
\nabla_\mu V^\mu = \frac{1}{\sqrt{{-g}}}\,\partial_\mu(\sqrt{{-g}}\,V^\mu)\,\,,
\end{equation}
together with \eq{solids-unit_vector}, we have 
\begin{equation}
\label{eq:appendix-A-4}
\delta\theta = \frac{1}{2}\theta g_{\mu\nu}\delta g^{\mu\nu} - \frac{1}{2}\nabla_\mu(O^\mu g_{ij}\delta g^{ij})\,\,. 
\end{equation}
Correspondingly, \eqsI{appendix-A-2} become 
\begin{subequations}
\label{eq:appendix-A-5}
\begin{align}
3\mpl^2H^2 &= \frac{\partial{\cal L}_0}{\partial\ln\theta} - {\cal L}_0\,\,, \label{eq:appendix-A-5-1} \\
{{-\frac{\dot{H}}{H^2}}} &= {\frac{X({\cal L}_{0,X} - \theta{\cal L}_{0,X\theta})}{({\cal L}_{0} - \theta{\cal L}_{0,\theta}) 
\big(1-\frac{3}{2}\frac{{\cal L}_{0,\theta\theta}}{M^2_{\rm P}}\big)}\,\,.} \label{eq:solid-13-2}
\end{align}
\end{subequations} 
where {we use a comma to denote partial derivatives with respect to \sm{X} and \sm{\theta}.} 
These relations are used in the same way as in Solid Inflation: one expands \sm{{\cal L}_0} in perturbations around an FLRW metric, 
and the derivatives with respect to \sm{X} and \sm{\theta} can be replaced by the expansion history using 
\eqsI{appendix-A-5}.\footnote{Higher derivatives with respect to these variables can be computed recursively. For 
example, at zeroth order in derivatives (i.e.~in Solid Inflation) one has 
\begin{equation}
\label{eq:appendix-A-6}
{\frac{X^2{\cal L}_{0,XX}}{X{\cal L}_{0,X}} = {-1}+\varepsilon-\frac{\eta}{2}\,\,.} 
\end{equation}} 
\item[Up to fourth order in derivatives] At this order we cannot get away anymore with just 
adding a dependence on \sm{\theta}. This is exemplified by the operator 
\begin{equation}
\label{eq:appendix-A-7}
(O^\rho\nabla_\rho{\cal K}_{\mu\nu})(O^\sigma\nabla_\sigma{\cal K}^{\mu\nu})\,\,,
\end{equation} 
whose background contains \sm{(O^\mu\nabla_\mu\theta)^2}. Still, it is enough to include a dependence on the operator 
\sm{\dot{\theta}\equiv O^\mu\nabla_\mu\theta} (and on this operator only) in \sm{{\cal L}_0}. 
\item[Higher orders] The same happens at higher orders. We always need to add a finite number of operators to \sm{{\cal L}_0} at any order in derivatives: 
these are essentially ``time derivatives'' of \sm{\theta}. The tadpole cancellation can be implemented in a similar way as the procedure leading to 
\eqsI{appendix-A-5}. 
\end{description} 

In this way we can write an ``EFT of Inflation'' for solids, with all the good 
properties of the one for broken time diffeomorphisms. More precisely, we write 
\begin{equation}
\label{eq:appendix-A-8}
S = S_{\rm EH} + {\underbrace{\int\dif^4x\,\sqrt{-g}\,{\cal L}_0(X,\theta,\dot{\theta},\dots)}_{
\hp{S_0\,}\equiv\,S_0}} + \text{EFT operators}\,\,, 
\end{equation} 
where the ``EFT operators'' start at second or higher order in perturbations around FLRW. 
One can turn on a time dependence of the coefficients of these operators by making them dependent on \sm{X}, 
similarly to how the coefficients in the EFT of Inflation are allowed to be dependent on time. 
In \sm{S_0} we keep a finite number of derivatives of \sm{\theta} depending on the order in derivatives at 
which we stop in the overall action. The only complication is that the background equations, which determine 
the form of \sm{S_0}, must be solved every time. 
This is only a slight complication given the very simple form of the functional derivatives of 
\sm{X,\theta,\dot{\theta},\dots} around a metric as symmetric as FLRW. 

It would be interesting to study how the additional operators \sm{\theta,\dot{\theta},\dots} present in 
\sm{S_0} affects the dynamics of the Goldstone modes. This is equivalent in studying the predictions of the 
``slow-roll action'' of the EFT of Inflation. Unlike what happens here, in the EFT of Inflation \sm{S_0} is fixed at all orders 
in derivatives: it is just \eq{EFTI-general_action} with \sm{c={-\mpl^2\dot{H}}} and \sm{\Lambda = \mpl^2(3H^2+\dot{H})}. 
We leave this to future work.

\section{Cubic graviton interaction in Solid Inflation}
\label{app:appendix-B}

\noindent Let us quickly show that the cubic graviton action at zeroth order in derivatives is not 
suppressed by \sm{\varepsilon}. Expanding \eq{solids-general_action} in perturbations, one finds 
(using the notation \sm{\partial F/\partial X = F_X}, \sm{\partial F/\partial Y = F_Y}, 
\sm{\partial F/\partial Z = F_Z} for compactness{, following \cite{Endlich:2012pz},} and dropping integral signs) 
\begin{subequations}
\label{eq:appendix-B-1}
\begin{align}
\frac{S_{\gamma\gamma}}{a^3} &\supset \bigg(\frac{1}{6}\frac{\partial F}{\partial\ln X} 
+ \frac{1}{9}F_Y + \frac{1}{9}F_Z\bigg)\gamma_{ij}\gamma_{ij}\,\,, \label{eq:appendix-B-1-1} \\
\frac{S_{\gamma\gamma\gamma}}{a^3} &\supset -\bigg(\frac{1}{18}\frac{\partial F}{\partial\ln X} 
+ \frac{1}{9}F_Y + \frac{4}{27}F_Z\bigg)\gamma_{ij}\gamma_{jk}\gamma_{ki}\,\,. \label{eq:appendix-B-1-2}
\end{align}
\end{subequations} 
Using \eq{solids-SR_condition} and the relation \sm{3\mpl^2H^2 = {-F}}, 
together with the fact that the propagation speeds \sm{c^2_{\rm L}} and \sm{c^2_{\rm T}} of the longitudinal and transverse part of \sm{\pi^i} are \cite{Endlich:2012pz} 
\begin{subequations}
\label{eq:appendix-B-2}
\begin{align}
c^2_{\rm L} &= \frac{1}{3} + \frac{8}{9}\frac{F_Y + F_Z}{X F_X}\,\,, \label{eq:appendix-B-2-1} \\
c^2_{\rm T} &= 1 + \frac{2}{3}\frac{F_Y + F_Z}{X F_X}\,\,, \label{eq:appendix-B-2-2}
\end{align}
\end{subequations} 
we see that \eq{appendix-B-1-1} is a small mass 
\begin{equation}
\label{eq:appendix-B-3}
m^2_\gamma = {-4\dot{H}c^2_{\rm T}}\,\,,
\end{equation}
while \eq{appendix-B-1-2} gives 
\begin{equation}
\label{eq:appendix-B-4}
\frac{S_{\gamma\gamma\gamma}}{\mpl^2 a^3} \supset {-\frac{1}{9}}\frac{F_Y}{F}H^2\gamma_{ij}\gamma_{jk}\gamma_{ki} 
+ \frac{1}{2} \varepsilon H^2c^2_{\rm L}\gamma_{ij}\gamma_{jk}\gamma_{ki}\,\,. 
\end{equation} 
The first term here reproduces Eq.~(A.6) of \cite{Endlich:2013jia}.

\section{{Equivalent of three-Ricci scalar for broken spatial diffeomorphisms}}
\label{app:appendix-C}

\noindent In this appendix we show how to construct an operator that reduces to the three-Ricci scalar if 
the graviton only is considered. We only need to recall the Gauss relation 
\begin{equation}
\label{eq:appendix-C-1}
h_\mu^{\hp{\mu}\alpha}h_\nu^{\hp{\nu}\beta}h_\gamma^{\hp{\gamma}\rho}h_\sigma^{\hp{\sigma}\lambda}R^\gamma_{\hp{\gamma}\lambda\alpha\beta} 
= {{^{(3)}}\!R^{\rho}_{\hp{\rho}\sigma\mu\nu}} + K_\mu^{\hp{\mu}\rho}K_{\nu\sigma} - K_\nu^{\hp{\nu}\rho}K_{\mu\sigma}\,\,,
\end{equation}
where our convention for the Riemann tensor is \sm{R^{\rho}_{\hp{\rho}\sigma\mu\nu}V^\sigma 
= [\nabla_\mu,\nabla_\nu]V^\rho}. Then, the object 
\begin{equation}
\label{eq:appendix-C-2}
{{^{(3)}}{\cal R}} = {\cal H}^{\lambda\beta}{\cal H}_\gamma^{\hp{\gamma}\alpha}R^\gamma_{\hp{\gamma}\lambda\alpha\beta} 
- {\cal K}^2 + {\cal K}_{\mu\nu}{\cal K}^{\mu\nu}\,\,, 
\end{equation} 
where \sm{{\cal H}_{\mu\nu} = g_{\mu\nu} + O_\mu O_\nu}, 
\sm{{\cal K}_{\mu\nu}} is defined by \eq{solids-extrinsic_curvature-A}, and ${\cal K} = {\cal H}_{\mu\nu}{\cal K}^{\mu\nu} = \theta$, 
reduces to \sm{\tr} if only tensor perturbations are considered.



\clearpage

\bibliographystyle{utphys}
\bibliography{refs}


\end{document}